\documentclass[english]{article}
\usepackage[T1]{fontenc}
\usepackage[latin9]{inputenc}
\usepackage{amstext}
\usepackage{setspace}
\makeatletter
\usepackage[numbers, sort&compress]{natbib}

\makeatother

\usepackage{babel}
\begin{document}

\title{Comment on (t, n) Threshold d-level Quantum Secret Sharing}

\author{Shih-Hung Kao and Tzonelih Hwang\thanks{Corresponding author\protect \\
hwangtl@csie.ncku.edu.tw\protect \\
Department of Computer Science and Information Engineering, National
Cheng Kung University, No. 1, University Rd., Tainan City, 701, Taiwan,
R.O.C.}}
\maketitle
\begin{abstract}
This comment points out a problem in Song et al.'s (t, n) threshold
quantum secret sharing {[}Scientific Reports, Vol. 7, No. 1 (2017),
pp. 6366{]}, indicating that the agent is unable to obtain the expected
information.
\end{abstract}

\paragraph{$ $}

In 2017, Song et al. {[}1{]} proposed an interesting way to perform
the summation of Lagrange interpolation formula in Shamir's secret
sharing (SS) by using quantum mechanics. Based on this method, the
agents in the QSS can obtain the boss's secret without announcing
any information. However, this study points out a calculation problem,
which indicates that the agent is unable to obtain the boss's secret
information without publishing any information.

\paragraph{Song et al.'s Secret Reconstruction Protocol.}

In Song et al.'s QSS, the boss uses Shamir's SS to generate the polynomial
$f\left(x\right)=a_{0}+a_{1}x+a_{2}x^{2}+...+a_{t-1}x^{t-1}$ and
$n$ distinct points of $\left(x_{i},\text{ }f\left(x_{i}\right)\right)$,
where $a_{0}$, $a_{t-1}\in Z_{d}$ and $i=1\sim n$. When $t$ Bobs
(the agents) want to recover the boss's secret, $a_{0}$, they will
execute the secret reconstruction protocol to compute the Lagrange
interpolation formula. Without loss of generality, assume that Bob$_{2}$
\textasciitilde{} Bob$_{t}$ help Bob$_{1}$ to obtain $a_{0}$. The
secret reconstruction protocol is as follows:
\begin{description}
\item [{Step1}] Bob$_{1}$ generates the d-level quantum state $\left|\varphi\right\rangle =\frac{1}{\sqrt{d}}\sum_{k=0}^{d-1}\left|kk...k\right\rangle _{12...t}$,
where the subscript 1 (2, 3, .., t) denotes the first (second, third,
..., t-th) qubit, respectively. Bob$_{1}$ sends the r-th qubit to
Bob$_{r}$ via the authenticated quantum channel, where $r=2\sim t$.
\item [{Step2}] Every participant computes $s_{r}=f\left(x_{r}\right)\prod_{1\leq j\leq t,j\neq r}\frac{x_{j}}{x_{j}-x_{r}}\text{ }\text{mod}\text{ }d$
according to Lagrange interpolation formula, where $r=1\sim t$. Then,
Bob$_{r}$ performs $U_{0,s_{r}}$ on their r-th qubit, where $r=1\sim t$
and $U_{0,s_{r}}$ is defined as:
\begin{equation}
U_{0,s_{r}}=\sum_{k=0}^{d-1}\omega^{s_{r}\cdot k}\left|k\right\rangle _{rr}\left\langle k\right|,
\end{equation}
where $\omega=e^{2\pi i/d}$. After these operations, the entire quantum
system will become: 
\begin{equation}
\begin{array}{lll}
\left|\varphi^{\prime}\right\rangle  & = & \frac{1}{\sqrt{d}}\sum_{k=0}^{d-1}\omega^{s_{1}\cdot k}\left|k\right\rangle _{1}\omega^{s_{2}\cdot k}\left|k\right\rangle _{2}...\omega^{s_{t}\cdot k}\left|k\right\rangle _{t}\\
 & = & \frac{1}{\sqrt{d}}\sum_{k=0}^{d-1}\omega^{\left(\sum_{r=1}^{t}s_{r}\right)\cdot k}\left|k\right\rangle _{1}\left|k\right\rangle _{2}...\left|k\right\rangle _{t}
\end{array}
\end{equation}
\item [{Step3}] Without receiving any information from the other agents,
Bob$_{1}$ can perform inverse quantum Fourier transform ($QFT^{-1}$)
on his first qubit and measures it with the basis $\left\{ \left|0\right\rangle ,\left|1\right\rangle ,...,\left|d-1\right\rangle \right\} $.
The measurement result, $a_{0}^{\prime}$, should be equal to the
boss's secret information, $a_{0}$.
\end{description}

\paragraph{The Problem.}

In Step 3, Song et al. claimed that when Bob$_{1}$ performs $QFT^{-1}$
on the first qubit, the quantum system will be
\begin{equation}
\begin{array}{lll}
QFT^{-1}\left(\frac{1}{\sqrt{d}}\sum_{k=0}^{d-1}\omega^{\left(\sum_{r=1}^{t}s_{r}\right)\cdot k}\left|k\right\rangle _{1}\right) & =\frac{1}{\sqrt{d}} & \sum_{k=0}^{d-1}QFT^{-1}\left(\omega^{\left(\sum_{r=1}^{t}s_{r}\right)\cdot k}\left|k\right\rangle _{1}\right)\\
 & = & \left|\sum_{r=1}^{t}s_{r}\text{ }\text{mod}\text{ }d\right\rangle _{1}\\
 & = & \left|a_{0}\text{ }\text{mod}\text{ }d\right\rangle _{1}.
\end{array}
\end{equation}

However, Bob$_{1}$'s qubit is actually entangled with the other participants'
qubits and the derivation should include the entire quantum system,
which should be written as follows:
\begin{equation}
\begin{array}{ll}
QFT^{-1}\otimes I\otimes...\otimes I\left(\frac{1}{\sqrt{d}}\sum_{k=0}^{d-1}\omega^{\left(\sum_{r=1}^{t}s_{r}\right)\cdot k}\left|k\right\rangle _{1}\left|k\right\rangle _{2}...\left|k\right\rangle _{t}\right) & =\\
\frac{1}{\sqrt{d}}\sum_{k=0}^{d-1}QFT^{-1}\left(\omega^{\left(\sum_{r=1}^{t}s_{r}\right)\cdot k}\left|k\right\rangle _{1}\right)\left|k\right\rangle _{2}...\left|k\right\rangle _{t}
\end{array}
\end{equation}
It can be seen that $QFT^{-1}\left(\omega^{\left(\sum_{r=1}^{t}s_{r}\right)\cdot k}\left|k\right\rangle _{1}\right)$
cannot be summed up together because $\left|k\right\rangle _{2}...\left|k\right\rangle _{t}$
are not equal to one another, where $k=0\sim d-1$. Hence, when Bob$_{1}$
measures his qubit, he cannot obtain the boss's secret information
$a_{0}$. In other words, Bob$_{1}$ cannot recover the boss's secret
without receiving any information from the other agents.

Let us take an example to explain the problem. Let $d=4$, $t=3$,
and $a_{0}=3$. After all Bobs' encoding, the quantum state is as
follows:
\begin{equation}
\begin{array}{lll}
\left|\varphi^{\prime}\right\rangle  & = & \frac{1}{2}\left(\omega^{3\cdot0}\left|0\right\rangle _{1}\left|0\right\rangle _{2}\left|0\right\rangle _{3}+\omega^{3\cdot1}\left|1\right\rangle _{1}\left|1\right\rangle _{2}\left|1\right\rangle _{3}+\right.\\
 &  & \left.\omega^{3\cdot2}\left|2\right\rangle _{1}\left|2\right\rangle _{2}\left|2\right\rangle _{3}+\omega^{3\cdot3}\left|3\right\rangle _{1}\left|3\right\rangle _{2}\left|3\right\rangle _{3}\right)
\end{array}
\end{equation}
When Bob$_{1}$ performs the $QFT^{-1}$ on the first qubit, the system
becomes:
\begin{equation}
\begin{array}{lll}
QFT^{-1}\otimes I\otimes I\left|\varphi^{\prime}\right\rangle  & = & \frac{1}{2}\left(QFT^{-1}\left(\omega^{3\cdot0}\left|0\right\rangle _{1}\right)\left|0\right\rangle _{2}\left|0\right\rangle _{3}+\right.\\
 &  & \left.QFT^{-1}\left(\omega^{3\cdot1}\left|1\right\rangle _{1}\right)\left|1\right\rangle _{2}\left|1\right\rangle _{3}+\right.\\
 &  & \left.QFT^{-1}\left(\omega^{3\cdot2}\left|2\right\rangle _{1}\right)\left|2\right\rangle _{2}\left|2\right\rangle _{3}+\right.\\
 &  & \left.QFT^{-1}\left(\omega^{3\cdot3}\left|3\right\rangle _{1}\right)\left|3\right\rangle _{2}\left|3\right\rangle _{3}\right)
\end{array}
\end{equation}
To expand Eq. (6), the $QFT^{-1}\left(\omega^{3\cdot0}\left|0\right\rangle _{1}\right)$
, $QFT^{-1}\left(\omega^{3\cdot1}\left|1\right\rangle _{1}\right)$,
$QFT^{-1}\left(\omega^{3\cdot2}\left|2\right\rangle _{1}\right)$,
and $QFT^{-1}\left(\omega^{3\cdot3}\left|3\right\rangle _{1}\right)$
in this equation can be written as follows, where $\omega^{x}=e^{2\pi ix/4}$.
\begin{equation}
\begin{array}{l}
QFT^{-1}\left(\omega^{3\cdot0}\left|0\right\rangle \right)=\frac{1}{2}\left(\left|0\right\rangle +\left|1\right\rangle +\left|2\right\rangle +\left|3\right\rangle \right)\\
QFT^{-1}\left(\omega^{3\cdot1}\left|1\right\rangle \right)=\frac{1}{2}\left(-i\left|0\right\rangle -\left|1\right\rangle +i\left|2\right\rangle +\left|3\right\rangle \right)\\
QFT^{-1}\left(\omega^{3\cdot2}\left|2\right\rangle \right)=\frac{1}{2}\left(-\left|0\right\rangle +\left|1\right\rangle -\left|2\right\rangle +\left|3\right\rangle \right)\\
QFT^{-1}\left(\omega^{3\cdot3}\left|3\right\rangle \right)=\frac{1}{2}\left(i\left|0\right\rangle -\left|1\right\rangle -i\left|2\right\rangle +\left|3\right\rangle \right)
\end{array}
\end{equation}
According to Eq. (7), Eq. (6) becomes:
\begin{equation}
QFT^{-1}\otimes I\otimes I\left|\varphi^{\prime}\right\rangle =\frac{1}{4}\left(\begin{array}{l}
\left(\left|0\right\rangle +\left|1\right\rangle +\left|2\right\rangle +\left|3\right\rangle \right)_{1}\left|00\right\rangle _{23}+\\
\left(-i\left|0\right\rangle -\left|1\right\rangle +i\left|2\right\rangle +\left|3\right\rangle \right)_{1}\left|11\right\rangle _{23}+\\
\left(-\left|0\right\rangle +\left|1\right\rangle -\left|2\right\rangle +\left|3\right\rangle \right)_{1}\left|22\right\rangle _{23}+\\
\left(i\left|0\right\rangle -\left|1\right\rangle -i\left|2\right\rangle +\left|3\right\rangle \right)_{1}\left|33\right\rangle _{23}
\end{array}\right)
\end{equation}

If the superposition states of the first qubit could be summed up
together, then the state of the first qubit would be $\left|3\right\rangle $,
which is the information that Bob$_{1}$ wants to obtain. However,
because the superposition states of the second and the third qubits
are different from the other superposition states (that is, $\left|00\right\rangle _{23}$,
$\left|11\right\rangle _{23}$, $\left|22\right\rangle _{23}$, and
$\left|33\right\rangle _{23}$), the superposition states of the first
qubit cannot be summed up and hence it is not equal to $\left|3\right\rangle $.

\section*{Acknowledgment}

This research is supported by the Ministry of Science and Technology,
Taiwan, R.O.C., under the Contract No. MOST 105-2221-E-006 -162 -MY2.

\section*{References}

{[}1{]} X.-L. Song, Y.-B. Liu, H.-Y. Deng, and Y.-G. Xiao, \textquotedblleft (t,
n) threshold d-level quantum secret sharing,\textquotedblright{} Scientific
Reports, vol. 7, no. 1, p. 6366, 2017.
\end{document}